\documentclass[a4paper,amsmath,amssymb,prl,twocolumn,superscriptaddress]{revtex4}
\usepackage{graphicx}
\usepackage{dcolumn}
\usepackage{bm}
\usepackage{latexsym}
\usepackage{epic,eepic}
\usepackage{subfigure}

\begin{document}

\author{Gavin D.\ McCullagh, Davide Cellai, Aonghus Lawlor and Kenneth
A.\ Dawson} \affiliation{Irish Centre for Colloid Science and Biomaterials, Department of Chemistry, University College Dublin,
Belfield, Dublin 4, Ireland}

\title{Finite energy extension of a lattice glass model}

\begin{abstract}
We extend a previously studied lattice model of particles with infinite
repulsions to the case of finite energy interactions.  The phase diagram
is studied using grand canonical Monte Carlo simulation.  Simulations of
dynamical phenomena are made using the canonical ensemble.  We find
interesting order-disorder transitions in the equilibrium phase diagram
and identify several anomalous regimes of diffusivity. These phenomena
may be relevant to the case of strong orientational bonding near
freezing.
\end{abstract}

\date{\today}
\maketitle

Dynamical arrest
\cite{lawlor2002prl,degregorio2004,biroli2002,jackle2002,garrahan2002,bouchaud2003}
is a ubiquitous phenomenon in nature, with many manifestations. Well
known examples include conventional glasses
\cite{debenedetti2001,angell1985,mezard2000}, but in modern condensed
matter studies the range of systems where the issues have become
relevant is remarkable. Thus, dynamical arrest is now believed to be a
useful description for gellation
\cite{dawson2001,zaccarelli2002,pham2002,eckert2002,mallamace2000,foffi2002},
`solidification', glassification
\citep{angell1985,stillinger1995,mezard1987,debenedetti2001}, jamming
\cite{liu1998,puertas2002,trappe2001} and the ergodic-to-non-ergodic
transition. Recent work tends to indicate that scientists increasingly
see these phenomena as manifestations of a single underlying set of
principles \cite{pham2002,dawson2002,liu1998,trappe2001,wolynes2002}.

Numerous attempts have been made to study simple models to develop ideas
of dynamical arrest
\cite{kirkpatrick1987a,kirkpatrick1987b,crisanti1995,jackle1994,sellitto2002,franz2002,cugliandolo1993,mezard2000,barrat2001,caglioti1997}.
Two approaches have been prominent in recent years, exemplified by the
Kob-Andersen kinetically constrained models \cite{kob1993,sellito2000},
and Hamiltonian-type models such as that introduced by Biroli and
M\'ezard (BM).  This model, introduced in Ref.~\cite{biroli2002},
permits particles of any given type $i$ to have a number of neighbors
less than or equal to a prescribed number, $c_{i}$. If $c_{i}$ is equal
to the coordination number of the lattice, we have a lattice gas model,
elsewhere the model is believed to be an attempt to represent packing
constraints \cite{biroli2002,dawson2002c,lawlor2002prl}. The
relationship between these two different kinetic models has never been
fully explored, though there has been a tendency to conclude that they
are quite similar in spirit.

In this paper we extend the concept of a Hamiltonian model to include
`soft' repulsions, rather than simply hard constraints. Thus, we now
stipulate that $c_{i}$ neighbors of a given type have zero energy and
additional particles have a repulsive energy $\epsilon(i)$ (in our case
chosen to be $V_0$).  Clearly, as we pass to the limit of infinite
repulsions we expect to recover the original BM model. In what will
follow the reader may assume that this is indeed the case, though that
limit is not explicitly shown. Here we focus on this generalization and
in particular two of its applications. Firstly, it gives additional
insights into the mechanisms involved in the dynamical arrest of such
models and secondly, such models may be applicable to systems with
oriented bonding effects.

In the work discussed below, we have calculated phase diagrams using the
grand canonical Monte Carlo method. The initial configurations of these
models are obtained by slowly annealing in the repulsions from a random
lattice gas configuration. Since we form crystals for a large portion of
the phase diagram, such annealing procedures may sometimes have to be
optimized over long periods of time.  Phase transition curves are
typically determined from system sizes of $L=12$, though in some cases
larger sizes have been used to study particular transitions. After
annealing, equilibration times of $\tau=2000$ MCS (Monte Carlo Sweeps)
and averaging times of $100,000$ MCS have been used. Again, some specific
areas of the phase diagram have been studied with averaging times of
several millions sweeps.

The dynamical studies, from which we calculate the mean squared distance
traveled ($\left<r^2\right>$) and consequently the single-particle
diffusion constant ($D$), are based on a single step canonical Monte
Carlo approach.  In this method, random movements are proposed and the
conventional Metropolis procedure used to choose successful moves.  We
have checked that both the known lattice gas
\cite{tahir-kheli-elliot1983} and BM limits are faithfully reproduced.

\begin{figure}
\includegraphics[height=0.95\linewidth,draft=false,angle=270]{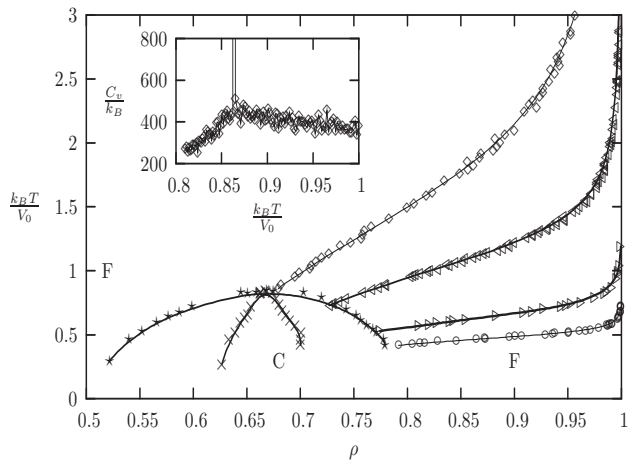}
\caption{The Phase Diagram of the $c=3$ system.  The points marked
{$\times$} bound the single-phase `crystalline' region C.  The points
marked {$\star$} bound the two-phase coexistence regions on either side
of C.  The region between C and F is therefore a coexistence region and
is determined from the discontinuities in the density for each constant
chemical potential.  However, the apex of the crystalline phase (that is
its high temperature limit) appears to lie on the line of fixed density
($\rho=\frac{2}{3}$) at which the crystal is perfectly packed.  The
nature of the transition there is uncertain.  To the left of this line,
the crystal (see Fig \ref{fig:subfig:holedefects}) has an excess of
vacancy defects and is in two-phase coexistence with a low density
isotropic fluid.  To the right of the line $\rho=\frac{2}{3}$, the
crystal (see Fig \ref{fig:subfig:partdefects}) has excess particle
defects and is in coexistence with a high density isotropic fluid.  The
points marked $\diamond,\triangleleft,\triangleright,\circ$ show lines
of constant chemical potential, $\beta\mu=6.0,9.0,15.0,20.0$
respectively.  Inset: $C_v/k_B$ vs $\frac{k_BT}{V_0}$ for constant
$\beta\mu=6.0$.} 
\label{fig1} 
\end{figure}

In Fig \ref{fig1} we present results for the simplest single component
system ($c=3$) model. The phase diagram for this system is quite
striking.  A `crystalline' region is bounded by first-order transitions
on the low density side by the gas and on the high density side by the
fluid. Apart from the special density $\rho=\frac{2}{3}$ where the
crystal is perfectly packed, the crystalline state contains many defects
and with the present level of simulation and theory it is not possible
to establish that it possesses true long ranged order.  Hence we have
referred to this state as `crystal'. The first-order phase-transition
curves from gas and fluid to the crystal terminate at high temperature
with a regime in which the crystal passes to the fluid.  The restricted
resolution of simulation prohibits us from deciding if this is a short
curve of weak first-order phase-transitions, or a single point of second
order (critical) transitions. We consider, on  quite general grounds
that either is possible.  In the inset to Fig \ref{fig1} we show the
heat capacity as one crosses (by lowering temperature) the transition
near the ``critical'' region ($\beta\mu=6.0$). The density difference
between gas or fluid appears to be nearly continuous, as illustrated in
the two-phase-regions of the phase diagram. If indeed the transition
were to be continuous, the topology of the phase diagram and form of the
heat capacity suggest that it would be a lambda-like transition
\cite{statmech_pathria_lambdatrans}.

\begin{figure}[!htp]
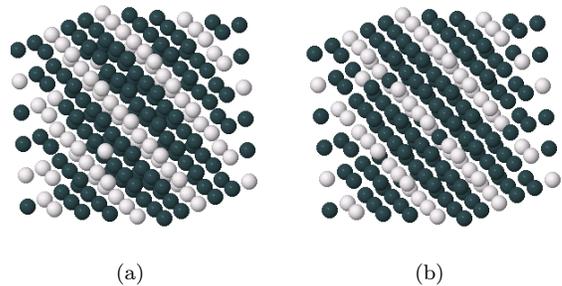

\begin{center}
\subfigure[]{%
    \label{fig:subfig:holedefects}%
    \includegraphics[scale=0.35,clip]{hole_defects2.epsi}%
}%
\qquad
\subfigure[]{%
   \label{fig:subfig:partdefects}%
   \includegraphics[scale=0.35,clip]{part_defects2.epsi}%
}
\caption{Samples of crystalline lattices at densities above and below
the ``critical'' density ($\rho_c=\frac{2}{3}$).  The black spheres
represent an occupied position on the lattice and the white ones
represent a vacancy.  Fig.~\ref{fig:subfig:holedefects} ($\rho=0.6424$)
shows hole defects within the occupied layers.
Fig.~\ref{fig:subfig:partdefects} ($\rho=0.6811$) shows particle defects
within the vacant layers.} 
\label{fig:defects}%
\end{center}
\end{figure}

The apex of the envelope of phase-transitions appears to occur precisely
at that density at which the crystal is perfectly packed
($\rho=\frac{2}{3}$). At higher or lower density (and temperatures lower
than this apex) the crystal has either an excess of particles or a
deficit of particles leading to vacancies. This point is illustrated in
Fig \ref{fig:defects}. The low density crystal
(\ref{fig:subfig:holedefects}) has extra vacancies in the
particle-filled layers, whereas the high density crystal
(\ref{fig:subfig:partdefects}) has extra particles in the `empty'
layers.  The defect density therefore vanishes on approach to this
special point, at density $\rho=\frac{2}{3}$.  Furthermore, at this
highest temperature the (free) energetic cost of these defects is
vanishing. It is therefore unclear whether the true order parameter of
the transition is the Fourier transform of the periodic density of the
crystal (as would be usual \cite{salinas1981, dawson1987}) or the defect
density. This is connected to the uncertainty in the nature of that
transition.

The equilibrium phase diagram is reminiscent of many structured fluids
composed of molecules that have strong directional bonding in the solid.
The `bonds' (in this case the repulsive energy between three of the
neighbors may be considered an effective attraction between the others)
are of finite energy, so that at low density or high temperature we have
a fluid, the crystal forming only when the effective bond energy becomes
sufficiently relevant.  As the packing density increases yet further,
the bonds are broken, the crystal re-melts to form a fluid.  In this
sense the model is an interesting simple example where ideas related to
such systems may be worked out.

We now study dynamical slowing across the phase diagram. For infinite
repulsions there is a transient slowing \cite{biroli2002, lawlor2002prl,
dawson2002c} that for some conditions or mixtures appears to be a
near-arrest transition. For the infinite repulsion case the onset of
this phenomenon always lies within the crystal regime and therefore
competes with crystallization. It is interesting to ask if there is any
true arrest for finite energy, or higher density. The answer we have
obtained for this model is, at first sight, a little surprising. The
transient phenomenon present at infinite repulsion is barely appreciable
for any finite repulsion and in the higher density re-melted regime
significant dynamical slowing (and indeed arrest) occurs only near the
fully filled lattice limit in which the system is close packed.  Thus,
even at very low temperatures, where the repulsions become very large,
providing the density is sufficiently high that the crystal melts there
is no arrest.

\begin{figure}
\includegraphics[height=0.95\linewidth,draft=false,angle=270]{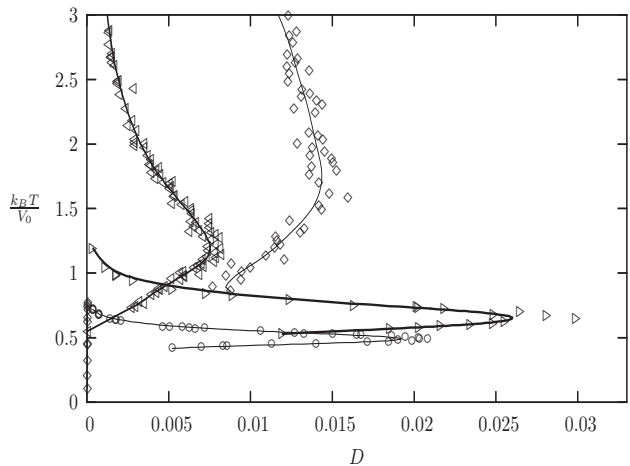}
\caption{Diffusion Constant ($D$) vs Temperature ($\frac{k_BT}{V_0}$) for
a set of constant chemical potentials, $\beta\mu=6.0,9.0,15.0,20.0$
marked $\diamond,\triangleleft,\triangleright,\circ$ respectively.  The peaks in the
$\beta\mu=15,20$ curves illustrate the anomalously high diffusion at
high density and low temperature.}
\label{fig3} 
\end{figure}

However, it is interesting that the diffusivity is non-monotonic in
density or temperature. We have studied the diffusion constants along a
series of iso-chemical potential curves ($\beta\mu=6.0, 9.0, 15.0,
20.0$).  These are given in Fig \ref{fig1} and the diffusion constants
along these curves are given in Fig \ref{fig3}.  In each case the
diffusion constant rises to a maximum as the temperature is lowered,
decreasing for yet lower temperatures. In the equivalent density
representation it is clear that there is a band of anomalously high
diffusivity in the high density fluid (re-melted) regime of the phase
diagram. The reasons for this are quite interesting and are related to
the issue of defects in the crystalline state. They merit a short
discussion.

At high density, as the fluid re-melts there develops a small population
of particles that cannot find a location where their three-bond limit is
satisfied. This is true even of the crystal, as noted earlier. Thus, in
the fluid, the density exceeds that limit at which every bond number is
satisfied. While it is true that these particles are unfavorable with
respect to energy, they are equally unfavorable no matter where they are
located in the system. They can therefore move with ease and become
`super-fast' diffusing particles. As the density increases the number of
such particles increases, conferring greater mobility to the system. At
yet higher density the repulsive energy is overcome, the system becomes
similar to the lattice gas and the diffusion constant decreases due to
the limited numbers of spaces into which particles can move. This
phenomenon is sufficiently general that it may be expected to occur in
the remelted regimes of structured fluids.

From a more general perspective we note that there is no significant
dynamical arrest phenomenon in the dense fluid, the diffusion constant
vanishing only at the close packed limit (lattice gas). We may conclude
that the blocking effects that lead to the transient near-arrest
phenomena in the infinite energy version of this model are strongly
associated with the crystalline region itself. That is, when the forces
are sufficient (and the density high enough) to destabilize the crystal,
dynamical arrest becomes less, not more likely.

In future it may be important to explore more carefully the mechanisms
of these hamiltonian and kinetic arrest models to see if indeed there
are more substantive differences that have hitherto not been recognized.

\acknowledgments{We acknowledge remarks from G.~Biroli, G.~Foffi,
M.~M\'ezard, F.~Sciortino, M.~Sellitto, B.~Widom and E.~Zaccarelli.
This paper is written within an EU-US research consortium, funded in
part by a grant from the Marie Curie program of the European Union,
(`Arrested Matter', contract number MRTN-CT-2003-504712).}

\bibliography{../bibtex/paper}

\end{document}